\documentclass[useAMS,usenatbib]{mn2e}
\usepackage{amsmath,amssymb,graphicx}

\title[Physical properties of WASP-44\,b]
  {Physical properties of the WASP-44 planetary system from simultaneous multi-colour photometry}
\author[L. Mancini et al.]
  {L.~Mancini,$^1$\thanks{mancini@mpia-hd.mpg.de}
  N.~Nikolov,$^1$ J.~Southworth,$^2$ G.~Chen,$^{1,3}$
  J. J.~Fortney,$^4$
  \newauthor 
  J.~Tregloan-Reed,$^2$ S.~Ciceri,$^1$
  R.~van Boekel,$^1$
  and Th. Henning$^1$ \\
  $^{1}$Max Planck Institute for Astronomy,
K\"{o}nigstuhl 17, 69117 -- Heidelberg, Germany \\
  $^2$Astrophysics Group, Keele University, Newcastle-under-Lyme, ST5 5BG, UK \\
  $^{3}$Purple Mountain Observatory \& Key Laboratory for Radio
Astronomy, 2 West Beijing Road, Nanjing 210008, China \\
$^{4}$Department of Astronomy \& Astrophysics, University of
California, Santa Cruz, CA 95064, USA}

\begin{document}

\date{Accepted. Received.}

\pagerange{\pageref{firstpage}--\pageref{lastpage}} \pubyear{2012}

\maketitle

\label{firstpage}

\begin{abstract}
We present ground-based broad-band photometry of two transits in
the WASP-44 planetary system obtained simultaneously through four
optical (Sloan $g^{\prime}$, $r^{\prime}$, $i^{\prime}$,
$z^{\prime}$) and three near-infrared (NIR; $J,\, H,\, K$)
filters. We achieved low scatters of 1--2 mmag per observation in
the optical bands with a cadence of $\approx48\,$s, but the
NIR-band light curves present much greater scatter. We also
observed another transit of WASP-44 b by using a Gunn-$r$ filter
and telescope defocussing, with a scatter of 0.37 mmag per point
and an observing cadence around $135\,$s. We used these data to
improve measurements of the time of mid-transit and the physical
properties of the system. In particular, we improved the radius
measurements of the star and planet by factors of 3 and 4,
respectively. We find that the radius of WASP-44\,b is $1.002 \pm
0.033 \pm 0.018 \,R_{\rm Jup}$ (statistical and systematic errors,
respectively), which is slightly smaller than previously thought
and differs from that expected for a core-free planet. In
addition, with the help of a synthetic spectrum, we investigated
the theoretically-predicted variation of the planetary radius as a
function of wavelength, covering the range $370$--$2440$\,nm. We
can rule out extreme variations at optical wavelengths, but
unfortunately our data are not precise enough (especially in the
NIR bands) to differentiate between the theoretical spectrum and a
radius which does not change with wavelength.
\end{abstract}


\section{Introduction}
\label{Sec_1}

The \emph{transit method} is not only an excellent technique to detect extrasolar planets, but also provide unrivalled access to their physical parameters. The geometry of a transiting extrasolar planet (TEP) system permits measurement of the mass, radius and density of a planet. These quantities require some input from stellar evolutionary theory, but the planetary surface gravity can be measured directly. A TEP system is therefore a potential \emph{horn of plenty} of physical information on planetary systems \citep{seager2000, sudarsky2000, brown2001, hubbard2001, seager2003, sudarsky2003, charbonneau2005, holman2005, winn2005, southworth2007}.

Currently, most of the TEPs discovered by ground-based transit surveys (e.g.\ \citealp{hellier2012,smalley2012,bakos2012,hartman2012,penev2012,bryan2012,beatty2012,siverd2012}) are close-in hot Jupiters, because the detection probability for such planets is much greater than for those which are smaller or on wider orbits. Hot Jupiters are also relatively straightforward to detect via the \emph{radial velocity method}, as they induce comparatively large velocity variations in their host stars. The existence of hot Jupiters was unexpected \citep{mayor1995}, and led to the birth of new formation and migration theories (see the reviews of \citealp{dangelo2010} and \citealp{lubow2010}). The bias of the ground-based surveys in favour of this class of planets is clear, whereas the {\it Kepler} satellite has already found hundreds of smaller and longer-period planet candidates from only 16 months of data \citep{borucki2011a, borucki2011b, batalha2012}.

Another fascinating opportunity offered by TEPs is the possibility to probe their atmospheres. Due to atomic and molecular absorption, the analysis of the spectrum of the parent stars during transit (primary eclipse) is an excellent way to probe their atmospheric composition. The existence of two different classes of hot Jupiters (pM and pL) has been suggested according to the incident stellar flux received from the parent star and to the expected amount of absorbing substances, such as gaseous titanium oxide (TiO) and vanadium oxide (VO), in their atmospheres \citep{fortney2008}. According to theoretical models, these two oxidized elements should be extremely strong absorbers for the pM class between 450\,nm and 700\,nm, causing such planets to have a hot ($\sim$$2000$ K) stratosphere. An observable variation of radius with wavelength, due to the opacities of these elements, is consequently expected in the optical bands \citep{burrows2007}. The radius of a pM planet should be $3\%$ lower at $350$--$400$\,nm versus $500$--$700$\,nm. The radius variations for the colder pL-class planets (incident flux $<$$10^{9}$\,erg\,s$^{-1}$\,cm$^{-2}$) are smaller, but a significant contribution to the opacity in the optical band is expected from Na\,{\small{I}} at $\sim$$590$\,nm, and K\,{\small{I}} at $\sim$$770$\,nm \citep{fortney2008, fortney2010, burrows2010}. Transmission spectroscopy is essential for detecting atmospheric absorbers, and is complementary to observations at secondary eclipse which can help to determine the day-side atmospheric temperature structure.

Accurate measurements of planet radii provide critical constraints for astrophysicists working on planet formation and evolution, because they provide an indication of the planet internal heat, structure and composition, as well as the heating mechanism for those which undergo strong tidal effects and stellar irradiation, allowing the discrimination between competing theories \citep{pollack1996,boss1997}. In particular, the most irradiated hot Jupiters are characterised by anomalously inflated radii for which the explanation remains unclear \citep{bodenheimer2003, gu2004, gaudi2005, fortney2007, jackson2008, dobbsdixon2008, batygin2009, ibgui2009, miller2009, ibgui2010, laughlin2011, miller2011, demory2011}.

Photometric and spectroscopic analyses of planet atmospheres started with the use of optical, near-infrared (NIR) and infrared (IR) instruments aboard the {\it Hubble} and {\it Spitzer} space telescopes \citep{charbonneau2002, vidalmadjar2003, vidalmadjar2004, richardson2006, tinetti2007, knutson2007a, knutson2007b, richardson2007, swain2008, pont2008, beaulieu2008, beaulieu2010, beaulieu2011, gillon2012} and led to the detection of Rayleigh scattering in a few hot Jupiter atmospheres, plus absorption lines associated with H$_2$O, Na, CH$_4$, TiO and VO (see above references but also \citealt{ballester2007, barman2007, sing2008a, sing2008b, lecavelier2008, desert2008, zahnle2009, spiegel2009, fossati2010, tinetti2010, desert2011, wood2011, gibson2011, crossfield2012b, crouzet2012}).

After a number of pioneering experiments with null results, ground-based spectroscopy also obtained some interesting results. Observations performed at the Hobby-Eberly and Subaru telescopes, in the spectral range $500$--$900$\,nm, detected Na absorption in the transmission spectrum of HD\,189733\,b \citep{redfield2008} and HD\,209458\,b \citep{snellen2008}. NIR spectroscopy of HD\,209458\,b at the VLT in the range $2291$--$2349$\,nm reported a significant wavelength shift in absorption lines from carbon monoxide in the planet's atmosphere \citep{snellen2010}.

High-resolution NIR spectroscopic measurements at the Keck telescopes over the wavelength range $2100$--$2400$\,nm, revealed an upper limit for CO absorption in the atmosphere of HD\,209458\,b \citep{deming2005}, and that the super-Earth GJ\,1214\,b is a H-dominated planet \citep{crossfield2011}. GJ\,1214\,b was also studied through multi-object spectroscopy from 0.61 to 0.85 $\mu$m, and in the $J$, $H$, and $K$ atmospheric windows by using VLT/FORS and Magellan/MMIRS, the data being consistent with a featureless transmission spectrum for the planet \citep{bean2011}.

A radius variation with wavelength was investigated for HD\,209458\,b by \citet{knutson2007a} using ten-colour HST photometry. Different analyses of these data have given conflicting results \citep{knutson2007a,barman2007,sing2008a,southworth2008}. An HST transmission spectrum of HD\,189733\,b covering 270--570\,nm showed a gradual increase of radius towards shorter wavelengths which was interpreted as a result of Rayleigh scattering from a high-altitude atmospheric haze \citep{sing2011a}. But observations covering 550--1050\,nm did not provide any indication of the expected Na\,{\small{I}} or K\,{\small{I}} features \citep{pont2008}, and those in the ranges $1082$--$1168$\,nm and $1514$--$1693$\,nm did not detect any variation in planetary radius \citep{gibson2012}.

Using the Gran Telescopio Canarias (GTC), \citet{sing2011b} claimed the first evidence for potassium in an extrasolar planet, from photometry of the XO-2 system in four narrow red-optical passbands, by detecting K\,{\small{I}} absorption at 766.5\,nm. This technique was also used by \citet{colon2012} to study HD\,80606\,b, who found an unexplained large ($\approx4.2\%$) change in the apparent planetary radius between the wavelengths 769.9 and 777.4\,nm. The differential spectrophotometry technique was used to obtain two transit light curves of XO-2\,b with GTC by \citet{sing2012}, who detected significant absorption in the planetary atmosphere in a 50-\AA\ bandpass centred on the Na\,{\small{I}} doublet. Instead, the presence of an Na-rich atmosphere in WASP-29\,b was recently ruled out by \citet{gibson2012b}, using Gemini-South GMOS transit spectrophotometry.

\citet{southworth2012b} recently presented a study of possible
radius variations of the hot Jupiter HAT-P-5\,b, based on
photometry obtained simultaneous in the $u$, $g$, $r$ and $I$
passbands. The authors detected a gradual increase of the radius
between $450$\,nm and $850$\,nm, plus a substantially larger
planetary radius at $350$\,nm. The latter phenomenon can be
explained by systematic errors in the $u$-band photometry, but is
also consistent with Rayleigh scattering as in the case of
HD\,189733\,b. A similar multi-band investigation of HAT-P-8
\citep{mancini2013} suggests the presence of strong optical
absorbers near the terminator of this TEP.

Here we focus our attention on the planetary system WASP-44, discovered by \citet{anderson2011}. The system consists of a 0.89\,$M_{\mathrm{Jup}}$ planet on a $2.42$-day orbit around an inactive G8\,V star ($V=12.9$, $\mathrm{[Fe/H]} = +0.06$). In this work we present the first photometric follow-up since its discovery was announced, covering seven optical/NIR passbands. We refine the physical properties of the system and attempt to probe for radius variations in these passbands.

Our paper is structured as follows. In \S~\ref{Sec_2} we describe the instruments used for the observations and give some details concerning the data reduction. In \S~\ref{Sec_3} we illustrate the analysis of the data which led to the refinement of the orbital period and the physical properties of the WASP-44 system. Then, we examine the variation of the planetary radius with wavelength. Finally, in \S~\ref{Sec_4} we summarise our results.

\section{Observations and data reduction}
\label{Sec_2}

Two transits of WASP-44\,b were recorded on 2011 October 2 and 6,
using the \textbf{G}amma \textbf{R}ay Burst \textbf{O}ptical and
\textbf{N}ear-Infrared \textbf{D}etector (GROND) instrument
mounted on the MPG\footnote{Max Planck Gesellschaft.}/ESO 2.2\,m
telescope at ESO La Silla, Chile. GROND is an imaging system
capable of simultaneous photometric observations in four optical
(identical to Sloan $g^{\prime}$, $r^{\prime}$, $i^{\prime}$,
$z^{\prime}$) and three NIR ($J,\, H,\, K$) passbands
\citep{greiner2008}. Each of the four optical channels is equipped
with a back-illuminated $2048 \times 2048$ E2V CCD, with a field
of view (FOV) of $5.4^{\prime} \times 5.4^{\prime}$ at a scale of
$0.158^{\prime\prime}/\rm{pixel}$. The three NIR channels use
$1024 \times 1024$ Rockwell HAWAII-1 arrays with a FOV of
$10^{\prime}\times 10^{\prime}$ at
$0.6^{\prime\prime}/\rm{pixel}$.

We applied a slight telescope defocus and obtained repeated integrations during both runs. Autoguiding was used to keep the stars on the same pixels. All images were digitised using the fast read-out mode ($\sim$$10$\,s) to improve the time sampling. On 2011/10/02 we monitored the flux of WASP-44 in nearly photometric conditions, and saw a monotonical seeing decline from $0.50^{\prime\prime}$ to $1.03^{\prime\prime}$ as airmass decreased from 1.08 to 1.05 and then rose to 1.54. Exposures of 20\,s were used in the optical channels and stacks of four 4\,s images were obtained in the NIR channels. On 2011/10/06 we experienced worse atmospheric conditions, and the airmass changed from 2.07 to 1.05 during our observations. Images with exposure times of 25\,s were obtained in the optical channels, and stacks of six 4\,s images in the NIR channels. For the optical frames we selected a fast read-out mode, which provided a 15\,s readout. Due to the necessary synchronization of the optical and NIR starting times for each exposure, the observing cadence resulted to be of $\approx 48\,$ and $\approx 25\,$s in the optical and NIR bands respectively.

Another full transit of WASP-44\,b was observed through a Gunn $r$ filter on the night of 2011/09/02 using the Wide Field Camera (WFC) on the 2.5\,m Isaac Newton Telescope (INT) at La Palma. WFC is an optical mosaic camera consisting of four thinned EEV 2k$\times$4k CCDs, with a plate scale of $0.33^{\prime\prime}$\,pixel$^{-1}$. The telescope was heavily defocussed so the point spread function (PSF) of the target and comparison stars had not more than $35\,000$ counts per pixel. We used the fast readout mode to reduce readout time down to about $15\,$s. An exposure time of $2\,$min per frame was used, and the resulting observing cadence was roughly $135\,$s. The night was photometric. We were not able to autoguide the telescope as the autoguider is incorporated into the WFC and so was also defocussed. A summary of the observational data is reported in Table\,\ref{Tab_1}.

\begin{table*} \centering
\begin{minipage}{180mm}
\caption{Log of the observations presented in this work. $N_{\mathrm{obs}}$ is the number of observations and ``Moon illum.'' is the fractional illumination of the Moon at the midpoint of the transit. The aperture sizes are the radii of the software apertures for the star, inner sky and outer sky, respectively. $\beta$ is the factor used to inflate the errorbars (see \S~\ref{Sec_3}). Transit \#1 was observed using the 2.5~m INT in La Palma, whereas transits \#2 and \#3 using the MPG/ESO 2.2~m telescope in La Silla.}
\label{Tab_1}
\setlength{\tabcolsep}{5pt}
\begin{tabular}{ccccccccccc} \hline
Transit  &  Date  &  Start/end Time  &  $N_{\mathrm{obs}}$  &  Exposure  &  Filter  &  Airmass  &  Moon    &  Aperture    &  Scatter  &  $\beta$  \\
         &        &  (UT)            &                      &  time (s)  &          &           &  illum.  &  sizes (px)  &  (mmag)   &           \\
\hline
1 & 2011 09 03 & 00:28 - 00:57 & 156 & 120 & Gunn  $r$           & $1.60\rightarrow1.32\rightarrow 2.18$ & $31\%$ & 25, 35, 50      & 0.37 & 1.15 \\
2 & 2011 10 02 & 03:21 - 07:35 & 317 &  20 & Sloan $g^{\prime}$  & $1.08\rightarrow1.05\rightarrow 1.54$ & $40\%$ & 16, 35, 55      & 1.23 & 1.07 \\
2 & 2011 10 02 & 03:21 - 07:35 & 317 &  20 & Sloan $r^{\prime}$  & $1.08\rightarrow1.05\rightarrow 1.54$ & $40\%$ & 17, 40, 60      & 1.08 & 1.00 \\
2 & 2011 10 02 & 03:21 - 07:35 & 317 &  20 & Sloan $i^{\prime}$  & $1.08\rightarrow1.05\rightarrow 1.54$ & $40\%$ & 20, 30, 45      & 1.34 & 1.00 \\
2 & 2011 10 02 & 03:21 - 07:35 & 317 &  20 & Sloan $z^{\prime}$  & $1.08\rightarrow1.05\rightarrow 1.54$ & $40\%$ & 12, 25, 40      & 1.79 & 1.10 \\
2 & 2011 10 02 & 03:21 - 07:35 & 631 &   3 & $J$                 & $1.08\rightarrow1.05\rightarrow 1.54$ & $40\%$ & 6.0, 15.0, 23.0 & 7.66 & 1.09 \\
2 & 2011 10 02 & 03:21 - 07:35 & 631 &   3 & $H$                 & $1.08\rightarrow1.05\rightarrow 1.54$ & $40\%$ & 2.5, 11.5, 21.0 & 8.11 & 1.08 \\
2 & 2011 10 02 & 03:21 - 07:35 & 631 &   3 & $K$                 & $1.08\rightarrow1.05\rightarrow 1.54$ & $40\%$ & 3.0, 11.0, 21.0 & 5.98 & 1.15 \\
3 & 2011 10 06 & 03:21 - 07:35 & 317 &  25 & Sloan $g^{\prime}$  & $2.07\rightarrow 1.05$                & $80\%$ & 19, 30, 50      & 1.89 & 1.06 \\
3 & 2011 10 06 & 03:21 - 07:35 & 313 &  25 & Sloan $r^{\prime}$  & $2.07\rightarrow 1.05$                & $80\%$ & 22, 40, 60      & 1.70 & 1.32 \\
3 & 2011 10 06 & 03:21 - 07:35 & 313 &  25 & Sloan $i^{\prime}$  & $2.07\rightarrow 1.05$                & $80\%$ & 20, 40, 60      & 1.79 & 1.35 \\
3 & 2011 10 06 & 03:21 - 07:35 & 313 &  25 & Sloan $z^{\prime}$  & $2.07\rightarrow 1.05$                & $80\%$ & 16, 40, 60      & 2.52 & 1.26 \\
3 & 2011 10 06 & 03:21 - 07:35 & 623 &   3 & $J$                 & $2.07\rightarrow 1.05$                & $80\%$ & 9.0, 17.0, 25.0 & 5.23 & 1.38 \\
3 & 2011 10 06 & 03:21 - 07:35 & 623 &   3 & $H$                 & $2.07\rightarrow 1.05$                & $80\%$ & 6.5, 13.5, 22.0 & 6.96 & 1.34 \\
3 & 2011 10 06 & 03:21 - 07:35 & 623 &   3 & $K$                 & $2.07\rightarrow 1.05$                & $80\%$ & 3.5,  4.5, 18.0 & 7.65 & 1.04 \\
\hline \end{tabular} \end{minipage} \end{table*}

Reduction of the optical frames was undertaken using standard methods. We created master bias and flat-field images by median-combining sets of bias images and sky flats, and used them to correct the science images. Aperture photometry was performed using the {\sc idl}\footnote{The acronym {\sc idl} stands for Interactive Data Language and is a trademark of ITT Visual Information Solutions. For further details see http://www.ittvis.com/ProductServices/IDL.aspx.}/{\sc astrolib}\footnote{http://idlastro.gsfc.nasa.gov/} implementation of {\sc daophot} \citep{stetson1987,southworth2009a}. The apertures were placed manually and shifted to account for pointing variations, which were measured by cross-correlating each image against a reference image. We experimented with wide range of aperture sizes and retained those which gave photometry with the lowest scatter compared to a fitted model. The times of observation were converted from UTC to BJD(TDB) using the IDL procedures of \citet{eastman2010}.

Differential photometry was obtained in each filter using between two and four comparison stars, two of which were enough bright to produce count rates comparable to those of WASP-44. All good comparison stars were combined into one ensemble by weighted flux summation. Slow variations in the apparent brightness of the reference stars, primarily attributable to atmospheric effects, were treated by fitting a polynomial to regions outside transit, whilst simultaneously optimising the weights of the comparison stars. Given the limited number of comparison stars in the small field of view of GROND, and the shape of the slow brightness variations, we used a second-order polynomial versus time. Uncertainties introduced by this procedure were considered in the modelling of the light curves, as described in the next section.

\begin{table} \centering
\caption{Excerpts of the light curves of WASP-44. The full dataset will be made available at the CDS.}
\label{Tab_2}
\begin{tabular}{@{}llcrr@{}} \hline
Telescope & Filter & BJD (TDB) & Diff.\ mag. & Error \\ \hline
INT      & Gunn $r$           & 2455807.576418 &  0.00036 & 0.00041 \\
INT      & Gunn $r$           & 2455807.579370 &  0.00009 & 0.00042 \\[3pt]
ESO 2.2m & Sloan $g^{\prime}$ & 2455836.647443 &  0.00126 & 0.00100 \\
ESO 2.2m & Sloan $g^{\prime}$ & 2455836.649992 &  0.00210 & 0.00123 \\[3pt]
ESO 2.2m & Sloan $r^{\prime}$ & 2455836.647443 & -0.00097 & 0.00088 \\
ESO 2.2m & Sloan $r^{\prime}$ & 2455836.649992 &  0.00079 & 0.00109 \\[3pt]
ESO 2.2m & Sloan $i^{\prime}$ & 2455836.647443 &  0.00096 & 0.00109 \\
ESO 2.2m & Sloan $i^{\prime}$ & 2455836.649992 &  0.00138 & 0.00134 \\[3pt]
ESO 2.2m & Sloan $z^{\prime}$ & 2455836.647443 & -0.00143 & 0.00142 \\
ESO 2.2m & Sloan $z^{\prime}$ & 2455836.649992 &  0.00100 & 0.00179 \\ \hline
\end{tabular} \end{table}

The NIR frames were also calibrated in a standard way, including dark subtraction, flat correction and sky subtraction. The master sky images used in sky subtraction were created from two sets of 20-position dithering sky measurements, one before the science observation and one after. We performed aperture photometry on the calibrated NIR images as well. The aperture locations were determined using {\sc idl/find}. Various combinations of aperture and annulus sizes were check to find the best photometry. We also carefully made ensembles of comparison stars, and chose the group which showed the least deviation from the target. After normalising the target light curve with the composite reference light curve, we decorrelated the data with position, seeing and airmass in order to remove the correlated red noise. We extracted the optimal NIR light curves for the two nights according to the $rms$ of O--C residuals and consistency of the transit depth between two nights. Details are given in Appendix \ref{Appendix_A}. The resulting photometry is given in Table\,2.

Although we have performed all the possible calibrations and
corrections, the scatter of the NIR data is much larger than that
of the optical data, and the NIR light curves are heavily
dominated by red noise. This problem is related to the adopted
observing strategy that unfortunately did not allow a good SNR to
be obtained\footnote{After these first observations, we changed
our observing strategy adopting the \textit{defocussing} technique
for GROND. Thanks to this approach, we can now obtain scatter
smaller than 2\,mmag per observation in the NIR bands and smaller
than 1\,mmag in the optical ones, without compromising the
sampling (Nikolov et al., in prep.).}.

Our analysis also included the dataset presented by \citet{anderson2011}, which was obtained through a Gunn $r$ filter and using EulerCam mounted on the 1.2\,m Euler-Swiss telescope at ESO La Silla.

\section{Analysis}
\label{Sec_3}

We have measured the physical properties of the WASP-44 planetary system following the methodology of the \emph{Homogeneous Studies} project \citep{southworth2008,southworth2009,southworth2010,southworth2011,southworth2012}. We refer the reader to those works for a detailed description of the approach.

Since the absolute values of the observational errors from our pipeline (which come ultimately from the {\sc aper} subroutine) were found to be underestimated, we adopt the standard practice of rescaling them for each dataset to give a reduced $\chi^{2}$ of $\chi_{\nu}^{2}=1$. Then, in order to account for time-correlated errors (i.e.\ red noise, which can significantly affect ground-based data) and derive more realistic uncertainties, we inflated the errorbars further by multiplying the data weights by a factor $\beta \geq 1$. The $\beta$ approach (e.g.\ \citealt{pont2006,winn2007}) is a widely used way to assess red noise (e.g.\ \citealt{gillon2006, winn2008, winn2009, gibson2008, nikolov2012, southworth2012a, southworth2012b}). The factor $\beta$ is a measurement of how close the data noise is to the Poisson approximation, and is found by binning the data and evaluating the ratio between the size of the residuals versus what would be expected if the data followed Poisson statistics. We evaluated values of the $\beta$ factor for each individual transit and for groups of 10 datapoints; they are reported in Table\,\ref{Tab_1}. For a more exhaustive discussion about rescaling errorbars, see \citet{andrae2010}.

\subsection{Period determination} %
\label{Sec_3.1}

\begin{figure*} \resizebox{\hsize}{!}{\includegraphics{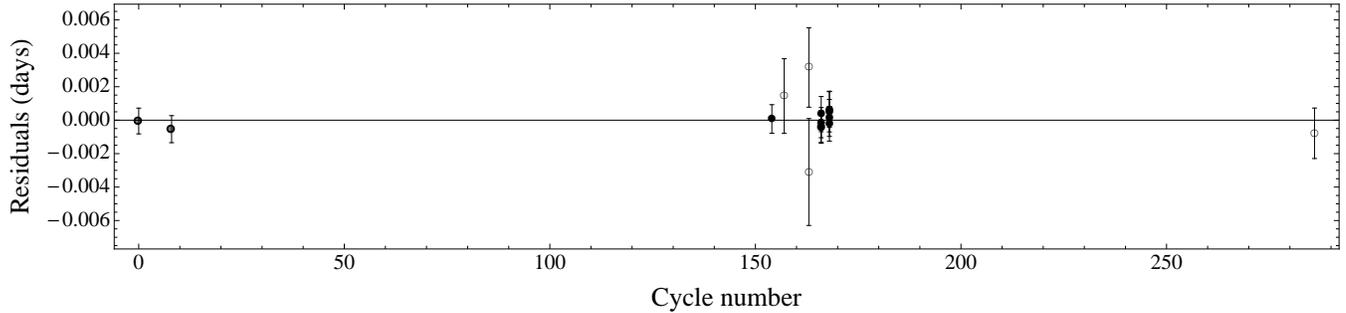}}
\caption{O-C diagram of the mid-transit times of WASP-44\,b versus a linear ephemeris. The timings in black are from this work, and in grey are from \citet{anderson2011}. The timings obtained by amateur astronomers are plotted using open circles. The uncertainties of our points have been rescaled (see text).}
\label{Fig_1}
\end{figure*}

As a first step, we fitted each light curve individually using the {\sc jktebop} code (see \S\,\ref{Sec_3.2}) in order to find the transit times. Their uncertainties were estimated from Monte Carlo simulations. To these we added two timings obtained by \citet{anderson2011} and four obtained by amateur astronomers, which are available on the TRESCA\footnote{The TRansiting ExoplanetS and CAndidates (TRESCA) website can be found at http://var2.astro.cz/EN /tresca/index.php} website (see Table\,\ref{Tab_3}). We excluded from this analysis our NIR data since they are much noisier than the optical ones. All timings were placed on BJD(TDB) time system. The resulting measurements of transit midpoints were fitted with a straight line to obtain a new orbital ephemeris:
\begin{equation} T_{0}=\mathrm{BJD(TDB)}2\,455\,434.37642(37)+2.4238133(23) \times E \end{equation}
where $E$ is the number of orbital cycles after the reference epoch (the midpoint of the first transit observed by \citealt{anderson2011}) and quantities in brackets denote the uncertainty in the final digit of the preceding number. The fit has $\chi_{\nu}^2=2.42$, so the uncertainties were increased to account for this. Despite this large $\chi_{\nu}^2$, we do not note any systematic deviation from the predicted transit times. A plot of the residuals around the fit is shown in Fig.\,\ref{Fig_1}. Since the number of known observed transits of this planet is still very low, no conclusions can be drawn regarding the existence of any third body in the system.

\begin{table} \centering
\caption{Central transit times of WASP-44 and their residuals versus the ephemeris derived in this work. TRESCA refers to the ``TRansiting ExoplanetS and CAndidates'' website.}
\label{Tab_3}
\setlength{\tabcolsep}{3pt}
\begin{tabular}{@{}lrrl@{}} \hline
Central transit time  & Cycle & Residual & Reference \\
BJD(TDB) -- $2400000$ & no.   & (JD)     &           \\ \hline
$55434.37637 \pm 0.00040$ & 0   & -0.00005 & \citet{anderson2011}           \\
$55453.76639 \pm 0.00042$ & 8   & -0.00054 & \citet{anderson2011}           \\
$55807.64374 \pm 0.00013$ & 154 &  0.00007 & This work (INT $r$)            \\
$55814.91655 \pm 0.00150$ & 157 &  0.00144 & Evans P. (TRESCA)              \\
$55829.45489 \pm 0.00245$ & 163 & -0.00310 & Lomoz F. (TRESCA)              \\
$55829.46151 \pm 0.00163$ & 163 &  0.00315 & Lomoz F. (TRESCA)              \\
$55836.72905 \pm 0.00020$ & 166 & -0.00038 & This work (GROND $g^{\prime}$) \\
$55836.72979 \pm 0.00030$ & 166 &  0.00036 & This work (GROND $r^{\prime}$) \\
$55836.72900 \pm 0.00020$ & 166 & -0.00043 & This work (GROND $i^{\prime}$) \\
$55836.72928 \pm 0.00015$ & 166 & -0.00015 & This work (GROND $z^{\prime}$) \\
$55841.57719 \pm 0.00035$ & 168 &  0.00014 & This work (GROND $g^{\prime}$) \\
$55841.57757 \pm 0.00046$ & 168 &  0.00052 & This work (GROND $r^{\prime}$) \\
$55841.57684 \pm 0.00028$ & 168 & -0.00021 & This work (GROND $i^{\prime}$) \\
$55841.57769 \pm 0.00031$ & 168 &  0.00064 & This work (GROND $z^{\prime}$) \\
$56127.58624 \pm 0.00048$ & 286 & -0.00078 & Sauer T. (TRESCA)              \\
\hline \end{tabular} \end{table}

\subsection{Light curve modelling} %
\label{Sec_3.2}

\begin{figure*} \resizebox{\hsize}{!}{\includegraphics{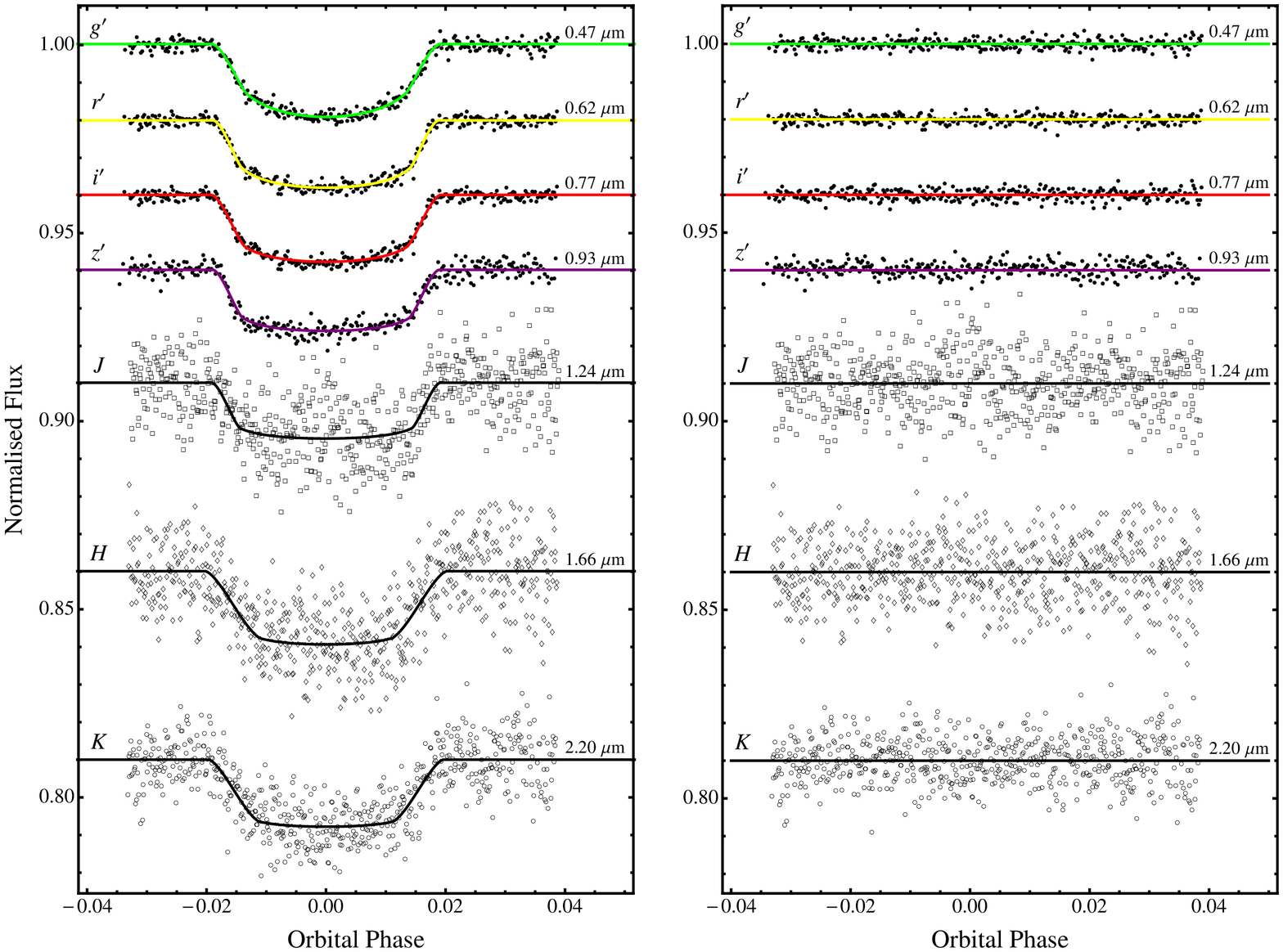}}
\caption{First set (2011/10/02) of GROND light curves of WASP-44 compared to the best \textsc{jktebop} fits using the quadratic LD law. The residuals of the fits are plotted at the base of the figure, offset from zero.}
\label{Fig_2}
\end{figure*}

\begin{figure*} \resizebox{\hsize}{!}{\includegraphics{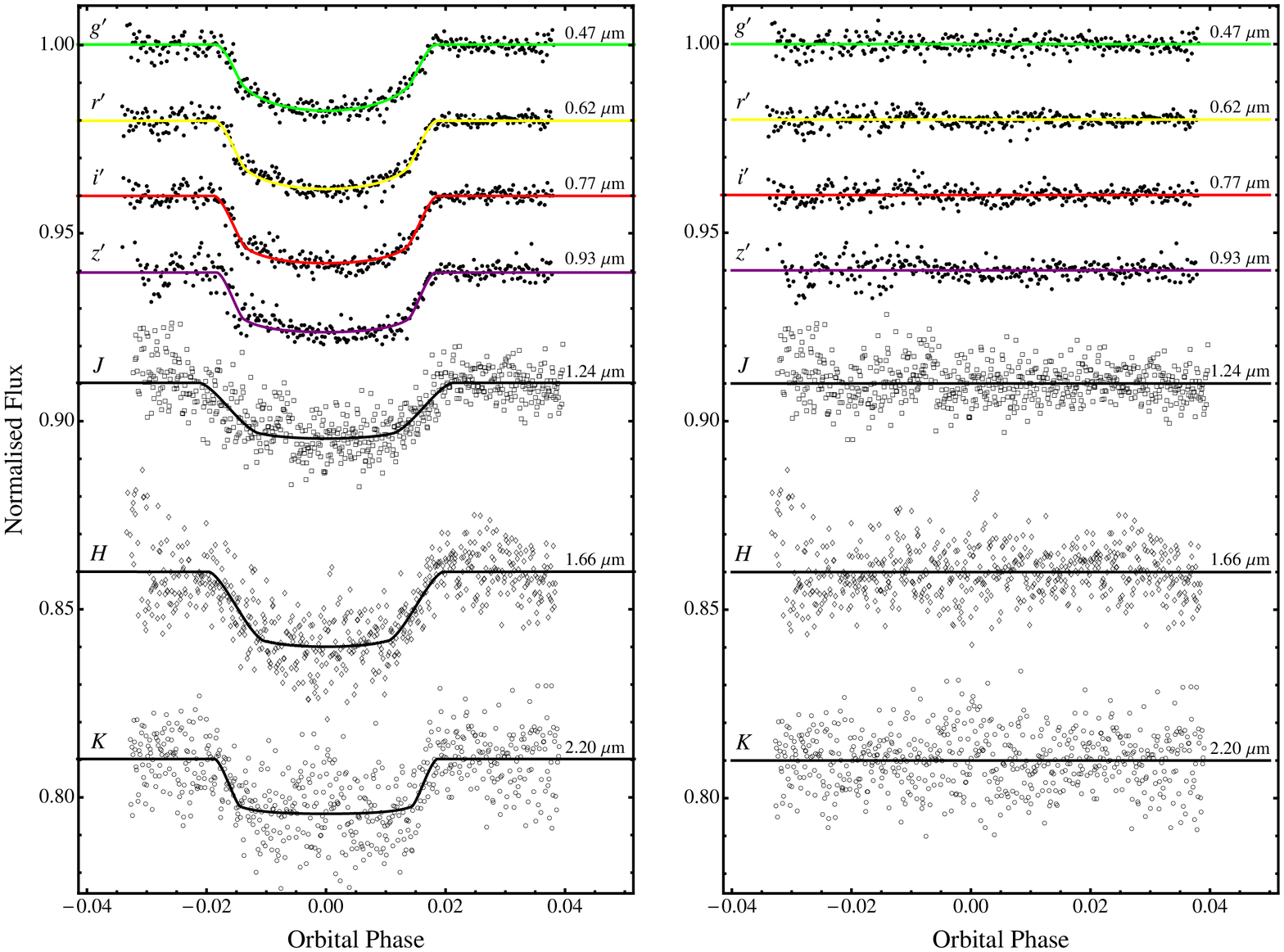}}
\caption{Second set (2011/10/06) of GROND light curves of WASP-44 compared to the best \textsc{jktebop} fits using the quadratic LD law. The residuals of the fits are plotted at the base of the figure, offset from zero.}
\label{Fig_3}
\end{figure*}

The light curves were modelled using the {\sc jktebop}\footnote{\textsc{jktebop} is written in FORTRAN77 and the source code is available at http://www.astro.keele.ac.uk/\textasciitilde jkt/} code. The primary fitted parameters were the sum and ratio of the fractional radii of the star and planet, $r_{\mathrm{A}}+r_{\mathrm{b}}$ and $k = r_{\mathrm{b}}/r_{\mathrm{A}}$, and the orbital inclination, $i$. The fractional radii of the components are defined as $r_{\mathrm{A}} = R_{\mathrm{A}}/a$ and $r_{\mathrm{b}} = R_{\mathrm{b}}/a$, where $a$ is the orbital semimajor axis, and $R_{\mathrm{A}}$ and $R_{\mathrm{b}}$ are the true radii of the two objects. Additional parameters of the fit included the light level outside transit and the midpoint of the transit. Limb darkening (LD) was imposed using a quadratic law and the corresponding coefficients fixed at values theoretically predicted using model atmospheres \citep{claret2004}. \citet{southworth2008} performed three different treatments of the LD coefficients: ($i$) both coefficients fixed to theoretical values; ($ii$) both coefficients fitted; ($iii$) the linear LD coefficient fitted and the nonlinear one fixed to its theoretically predicted value. While the second treatment is possible only when the data are of high quality, the use of the third alternative has a negligible impact on the final results. Uncertainties were calculated using both Monte Carlo simulations and a residual-permutation algorithm \citep{southworth2008}, and the larger of the two values was retained for each output quantity. The orbital eccentricity was fixed to zero \citep{anderson2011}. All the datasets were solved individually. As in the previous section, due to their large scatter (see Table\,\ref{Tab_1}), we did not use the NIR data for estimating the physical properties of the planetary system.

\begin{table*} \centering
\caption{Parameters of the fits to the ten light curves of WASP-44. The final parameters are the weighted mean of the result for the GROND, INT and Euler light curves.}
\label{Tab_4}
\begin{tabular}{@{}lccccc@{}} \hline
Source & $r_{\mathrm{A}}+r_{\mathrm{b}}$ & $k$ & $i^{\circ}$ & $r_{\mathrm{A}}$ & $r_{\mathrm{b}}$ \\ \hline
GROND $g^{\prime}-$band \#1 & $0.1368 \pm  0.0056$ & $0.1228 \pm 0.0023$ & $86.2 \pm 0.5$ & $0.1218 \pm 0.0020$ & $0.01496 \pm 0.00086$ \\
GROND $g^{\prime}-$band \#2 & $0.139  \pm  0.010$  & $0.1188 \pm 0.0028$ & $86.2 \pm 1.0$ & $0.1240 \pm 0.0090$ & $0.0147  \pm 0.0014$  \\
GROND $r^{\prime}-$band \#1 & $0.1268 \pm  0.0045$ & $0.1175 \pm 0.0014$ & $87.3 \pm 0.6$ & $0.1135 \pm 0.0039$ & $0.01333 \pm 0.00060$ \\
GROND $r^{\prime}-$band \#2 & $0.1284 \pm  0.0080$ & $0.1190 \pm 0.0029$ & $86.9 \pm 0.8$ & $0.1147 \pm 0.0068$ & $0.0136  \pm 0.0011$  \\
GROND $i^{\prime}-$band \#1 & $0.1363 \pm  0.0051$ & $0.1205 \pm 0.0016$ & $86.3 \pm 0.5$ & $0.1216 \pm 0.0044$ & $0.0147  \pm 0.00068$ \\
GROND $i^{\prime}-$band \#2 & $0.1303 \pm  0.0087$ & $0.1204 \pm 0.0029$ & $86.7 \pm 0.9$ & $0.1163 \pm 0.0076$ & $0.0140  \pm 0.0011$  \\
GROND $z^{\prime}-$band \#1 & $0.1306 \pm  0.0075$ & $0.1153 \pm 0.0023$ & $86.8 \pm 0.8$ & $0.1171 \pm 0.0065$ & $0.01349 \pm 0.00095$ \\
GROND $z^{\prime}-$band \#2 & $0.123  \pm  0.010$  & $0.1134 \pm 0.0034$ & $87.4 \pm 1.5$ & $0.1102 \pm 0.0089$ & $0.0125  \pm 0.0013$  \\
INT   $r-$band              & $0.1285 \pm  0.0029$ & $0.1196 \pm 0.0013$ & $86.6 \pm 0.3$ & $0.1148 \pm 0.0026$ & $0.01373 \pm 0.00042$ \\
Euler $r-$band              & $0.141  \pm  0.011$  & $0.1197 \pm 0.0039$ & $85.9 \pm 1.0$ & $0.1261 \pm 0.0100$ & $0.0151  \pm 0.0016$  \\ \hline
Final results               & & & $\mathbf{86.59 \pm 0.18}$ & $\mathbf{0.1168 \pm  0.0016}$ & $\mathbf{0.0139 \pm 0.0002}$ \\ \hline
\citet{anderson2011}        & $0.1398$ & $0.1260 \pm 0.0030$ & $86.02^{+1.11}_{-0.86}$ & $0.1242 \pm 0.0102$ & $0.0157$    \\ \hline
\end{tabular} \end{table*}

\begin{figure} \resizebox{\hsize}{!}{\includegraphics{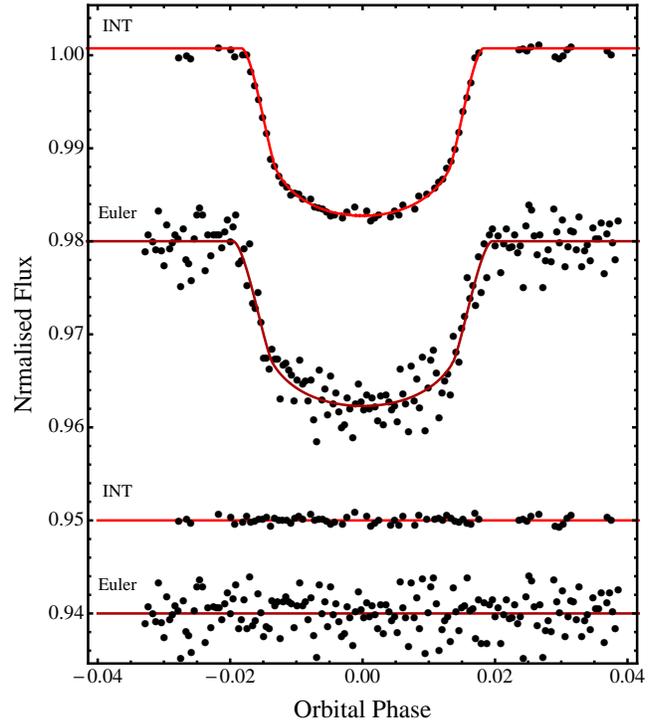}}
\caption{As for Fig.\,\ref{Fig_2} but for the $r-$filter INT and Euler \citep{anderson2011} light curves of WASP-44.}
\label{Fig_4}
\end{figure}

Representative photometric parameters were obtained for each dataset and are given in Table\,\ref{Tab_4}. The corresponding best fits are shown in Fig.\,\ref{Fig_2} and \ref{Fig_3} for the GROND data and in Fig.\,\ref{Fig_4} for the INT and Euler data. The final photometric parameters are the weighted mean of the values for each dataset. The agreement between light curves is excellent: the $\chi_\nu^2$ value of the agreement of the individual parameters with respect to the weighted mean is smaller than 0.75 for all except $k$, where we found a modestly larger $\chi_{\nu}^2$ of $1.64$. We have found this situation to be common during our work on the {\it Homogeneous Studies} papers. Table\,\ref{Tab_4} also shows a comparison with the results from \citet{anderson2011}, which are in good agreement with ours but have much larger errorbars.

\subsection{Physical properties of the WASP-44 system} %
\label{Sec_3.3}

Following the approach described in \citet{southworth2009}, the estimation of the physical properties of the WASP-44 system was performed making use of standard formula (e.g.\ \citealt{hilditch2001}). We used the photometric parameters measured in \S\,\ref{Sec_3.2}, and the velocity amplitude, effective temperature, and metallicity of the star ($K_{\mathrm{A}}=138.8 \pm 9.0$\,km\,s$^{-1}$, $T_{\mathrm{eff}}=5410 \pm 150$\,K, [Fe/H]$= +0.06 \pm 10$) measured by \citet{anderson2011}. We interpolated within tabulations from theoretical stellar models to find the best agreement between the observed and model-predicted $T_{\mathrm{eff}}$, and the measured $r_{\mathrm{A}}$ and calculated $R_{\mathrm{A}}/a$. This yielded the best-fitting mass, radius, surface gravity and mean density of the star ($M_{\mathrm{A}}$, $R_{\mathrm{A}}$, $\log g_{\mathrm{A}}$ and $\rho_{\mathrm{A}}$) and of the planet ($M_{\mathrm{b}}$, $R_{\mathrm{b}}$, $g_{\mathrm{b}}$ and $\rho_{\mathrm{b}}$). We also calculated the orbital semi-major axis ($a$), planetary equilibrium temperature ($T_{\rm eq}$), Safranov number ($\Theta$), and the evolutionary age of the star.

\begin{table*} \centering
\caption{Derived physical properties of WASP-44 from using each of five different theoretical stellar models. In each case $g_{\rm b} = 21.5 \pm 1.6$\,m\,s$^{-2}$, $\rho_{\rm A} = 1.414 \pm 0.058$\,$\rho_\odot$ and $T_{\rm eq} = 1304 \pm 37$\,K.}
\label{Tab_5}
\begin{tabular}{@{}lccccc@{}} \hline
& This work & This work & This work & This work & This work \\
& (Claret models) & (Y$^2$ models) & (Teramo models) & (VRSS models) & (DSEP models) \\\hline
$K_{\mathrm{b}}$ (km\,s$^{-1}$)             & $156.1   \pm 4.1$     & $151.7   \pm 1.0$     & $152.6   \pm 4.2$     & $152.7   \pm 4.1$     & $153.7   \pm 3.2$     \\
$M_{\mathrm{A}}$ ($M_{\sun}$)               & $0.968   \pm 0.077$   & $0.888   \pm 0.018$   & $0.904   \pm 0.074$   & $0.907   \pm 0.073$   & $0.924   \pm 0.058$   \\
$R_{\mathrm{A}}$ ($R_{\sun}$)               & $0.881   \pm 0.025$   & $0.856   \pm 0.016$   & $0.862   \pm 0.024$   & $0.862   \pm 0.024$   & $0.865   \pm 0.020$   \\
$\log g_{\mathrm{A}}$ (cgs)                 & $4.534   \pm 0.018$   & $4.521   \pm 0.010$   & $4.524   \pm 0.018$   & $4.524   \pm 0.018$   & $4.530   \pm 0.016$   \\
$M_{\mathrm{b}}$ ($M_{\mathrm{jup}}$)       & $0.901   \pm 0.075$   & $0.851   \pm 0.056$   & $0.861   \pm 0.073$   & $0.863   \pm 0.073$   & $0.874   \pm 0.067$   \\
$R_{\mathrm{b}}$ ($R_{\mathrm{jup}}$)       & $1.020   \pm 0.033$   & $0.992   \pm 0.019$   & $0.997   \pm 0.033$   & $0.998   \pm 0.032$   & $0.994   \pm 0.027$   \\
$\rho_{\mathrm{b}}$ ($\rho_{\mathrm{jup}}$) & $0.793   \pm 0.070$   & $0.816   \pm 0.069$   & $0.812   \pm 0.072$   & $0.811   \pm 0.072$   & $0.831   \pm 0.071$   \\
$\Theta$                                    & $0.0640  \pm 0.0046$  & $0.0658  \pm 0.0045$  & $0.0654  \pm 0.0048$  & $0.0654  \pm 0.0047$  & $0.0656  \pm 0.0046$  \\
$a$ (AU)                                    & $0.03508 \pm 0.00093$ & $0.03409 \pm 0.00023$ & $0.03429 \pm 0.00093$ & $0.03432 \pm 0.00092$ & $0.03454 \pm 0.00072$ \\
\hline \end{tabular} \end{table*}

The uncertainties in the input parameters were propagated into the output physical properties using a perturbation analysis \citep{southworth2005}. Systematic errors were assessed by comparing results from five different sets of theoretical models (see \citealt{southworth2010}). The sets of physical properties found using each set of stellar models is shown in Table\,\ref{Tab_5} and the final physical properties of the WASP-44 system are given in Table\,\ref{Tab_6}. The results obtained by \citet{anderson2011} are less precise but are in good agreement with our own. In particular, we improved the measurement precisions of the radii of both the planet and the parent star, by factors of 4 and 3 respectively. The radius of WASP-44\,b that we found ($1.002 \pm 0.033 \pm 0.018 \,R_{\rm Jup}$) is smaller than that measured in the discovery paper and does not match the predicted radii of coreless hot-Jupiter planets as estimated by \citet{fortney2007}. According to their tables\footnote{The tables, interpolated from the giant planet thermal evolution models described in \citet{fortney2007}, are available at http://www.ucolick.org/\textasciitilde jfortney/models.htm}, a $0.875\, M_{\mathrm{J}}$ planet with $50 \, M_{\mathrm{Earth}}$ core at 0.045 AU from the Sun (age 3.16 Gyr), is $0.991 \, R_{\mathrm{Jup}}$. On the contrary, the expected radius for a core-free planet with the same properties is $1.119  \, R_{\mathrm{Jup}}$.

\begin{table} \centering
\caption{Final physical properties of the WASP-44 system. The first errorbar for each parameter is the statistical error, which stems from the measured spectroscopic and photometric parameters. The second errorbar is the systematic error arising from the use of theoretical stellar models, and is given only for those parameters which have a dependence on stellar theory. The results from \citet{anderson2011} are included for comparison.}
\label{Tab_6}
\setlength{\tabcolsep}{2pt}
\begin{tabular}{@{}lcc@{}} \hline
& This work (final) & \citet{anderson2011} \\ \hline
$M_{\mathrm{A}}$ ($M_{\sun}$)               & $0.917   \pm 0.077   \pm 0.051$    & $0.951 \pm 0.034$         \\
$R_{\mathrm{A}}$ ($R_{\sun}$)               & $0.865   \pm 0.025   \pm 0.016$    & $0.927^{+0.068}_{-0.074}$ \\
$\log g_{\mathrm{A}}$ (cgs)                 & $4.526   \pm 0.018   \pm 0.008$    & $4.481^{+0.068}_{-0.057}$ \\
$\rho_{\mathrm{A}}$ ($\rho_{\sun}$)         & $1.414   \pm 0.058$                & $1.19^{+0.32}_{-0.22}$    \\
$M_{\mathrm{b}}$ ($M_{\mathrm{jup}}$)       & $0.869   \pm 0.075   \pm 0.032$    & $0.889 \pm 0.062$         \\
$R_{\mathrm{b}}$ ($R_{\mathrm{jup}}$)       & $1.002   \pm 0.033   \pm 0.018$    & $1.14 \pm 0.11$           \\
$g_{\mathrm{b}}$ ($\mathrm{ms^{-2}}$)       & $21.5    \pm 1.6$                  & $15.7_{-3.0}^{+3.4}$      \\
$\rho_{\mathrm{b}}$ ($\rho_{\mathrm{jup}}$) & $0.808   \pm 0.072   \pm 0.015$    & $0.61^{+0.23}_{-0.15}$    \\
$T_{\mathrm{eq}}$ ($\mathrm{K}$)            & $1304    \pm 37$                   & $1343 \pm 64$             \\
$\Theta$                                    & $0.0652  \pm 0.0048  \pm 0.0012$   & $-$                       \\
$a$ (AU)                                    & $0.03445 \pm 0.00093 \pm 0.00063$  & $0.03473 \pm 0.00041$     \\
Age (Gyr)                                   & $4.1_{-6.0\,-2.4}^{+5.9\,+4.1}$    & $-$                       \\
\hline \end{tabular} \end{table}

\subsection{Variation of planetary radius with wavelength} %
\label{Sec_3.4}

As an additional possibility for the GROND data, we made an attempt to investigate the possible variation of the radius of WASP-44\,b with wavelength. This is quite a difficult task because the relative faintness of the host star ($V \sim 12.9$) makes this system less than ideal for optical photometric studies, especially from the ground. Moreover, as already mentioned, our \emph{in-focus} observing strategy has proved to be ill-suited to obtaining high-precision photometry, particularly in the NIR bands. Finally, by using broad-band filters we are forced to average our measurements of the planet radius in each band over a quite larger range of the wavelength (the best case in the $i^{\prime}$ filter where its width is ``only'' 100 nm). For these reasons, the results of this section should be used with care.

The GROND instrument was conceived only for the follow-up of gamma-ray bursts, and was not designed to allow filter changes for individual observing sequences. We were therefore unable to use filters with narrow passbands covering specific wavelength ranges. We note that the four optical bands of GROND were already used by \citet{demooij2012} to investigate the atmosphere of GJ\,1214\,b, a 6.55\,M$_\oplus$ transiting planet. Here we try for the first time to use all the seven bands, exploiting the full potential of GROND.

\begin{figure*} \resizebox{\hsize}{!}{\includegraphics{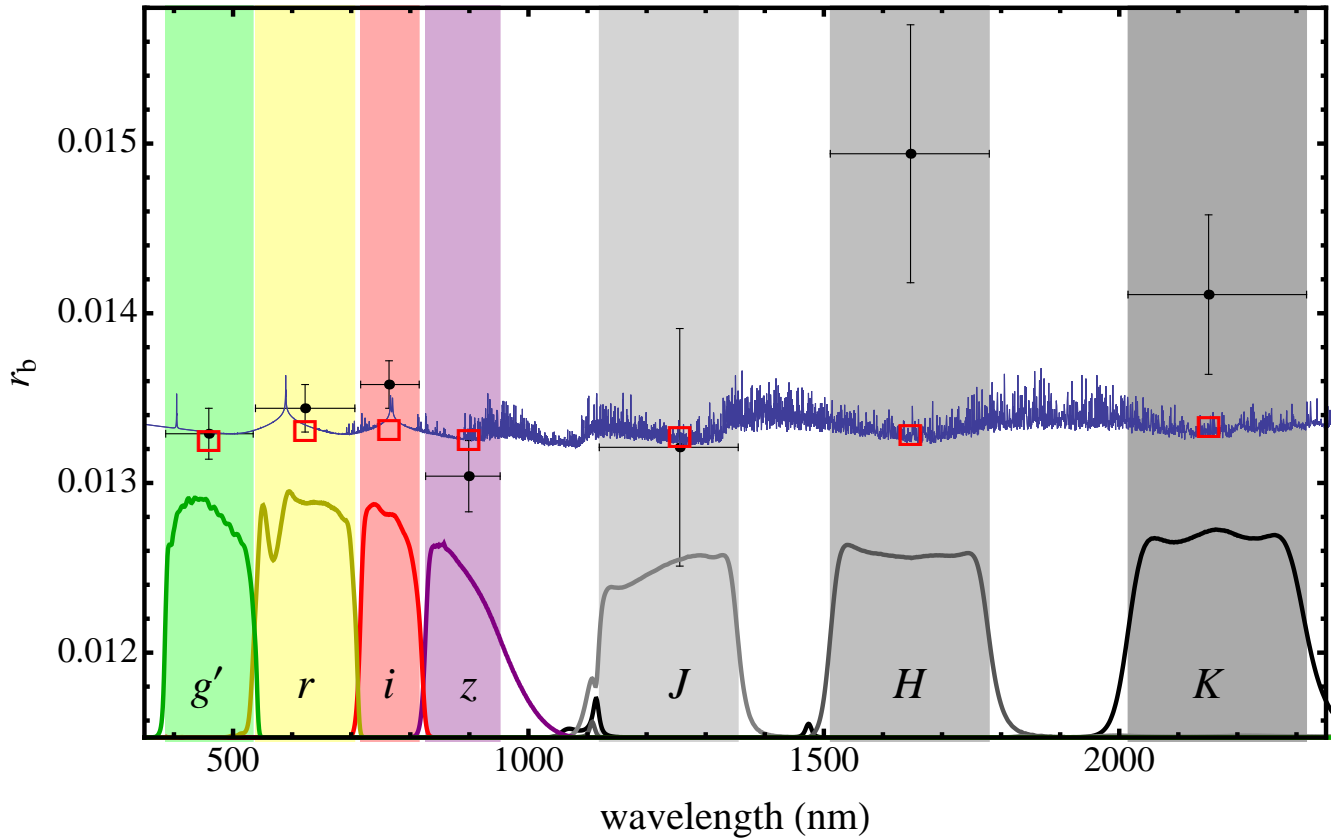}}%
\caption{Variation of the fractional planetary radius $r_{\mathrm{b}}=R_{\mathrm{b}}/a$ with wavelength. The points shown in the plot are from the MPG/ESO 2.2\,m telescope. The vertical bars represent the errors in the measurements and the horizontal bars show the full widths at half maximum transmission of the passbands used. The solid, blue continuous line is the calculated synthetic spectrum based on a theoretical model of the atmosphere of WASP-44\,b. Red boxes indicate the predicted values for this model integrated over the passbands of the observations. Transmission curves of the GROND filters are shown at the bottom of the figure.}
\label{Fig_5}
\end{figure*}

Following the same strategy used by \citet{southworth2012b} and \citet{mancini2013}, we proceeded as follows. The two GROND datasets were combined by phase and according to passband, and then fitted with all parameters fixed to the final values given in Table\,\ref{Tab_4}, with the exception of $k$. The LD coefficients were fixed to theoretical values. The errors were estimated by a residual-permutation algorithm \citep{southworth2008}. This approach removes sources of uncertainty common to all datasets, allowing us to maximise the accuracy of estimations of the fractional planetary radius $r_{\mathrm{b}}=R_{\mathrm{b}}/a$ as a function of wavelength and with relative errorbars only. The results are displayed in Fig.\,\ref{Fig_5}, where the points show the data, the vertical bars represent the relative errors in the measurements and the horizontal bars indicate the full widths at half maximum transmission of the passbands used. The trasmission curves are also reported for completeness. As expected, the uncertainties in the NIR bands are larger, due to the larger scatter and systematic features in the light curves.

Inspection of Fig.\,\ref{Fig_5} shows that our measurements are
unfortunately not sufficiently accurate to claim a clear variation
of $r_{\mathrm{b}}$ along the seven passbands. Using the planetary
system values reported in Table\,\ref{Tab_6}, we computed a
one-dimensional model atmosphere of WASP-44\,b, using the
atmosphere code described in \citet{Fortney05,fortney2008}. The
fully non-gray model uses the chemical equilibrium abundances of
\citet{Lodders02} and the opacity database described in
\citet{fr08}. The atmospheric pressure-temperature profile
simulates planet-wide average conditions. We furthermore computed
the transmission spectrum of the model using the methods described
in \citet{fortney2010}.

\begin{figure*} \resizebox{\hsize}{!}{\includegraphics{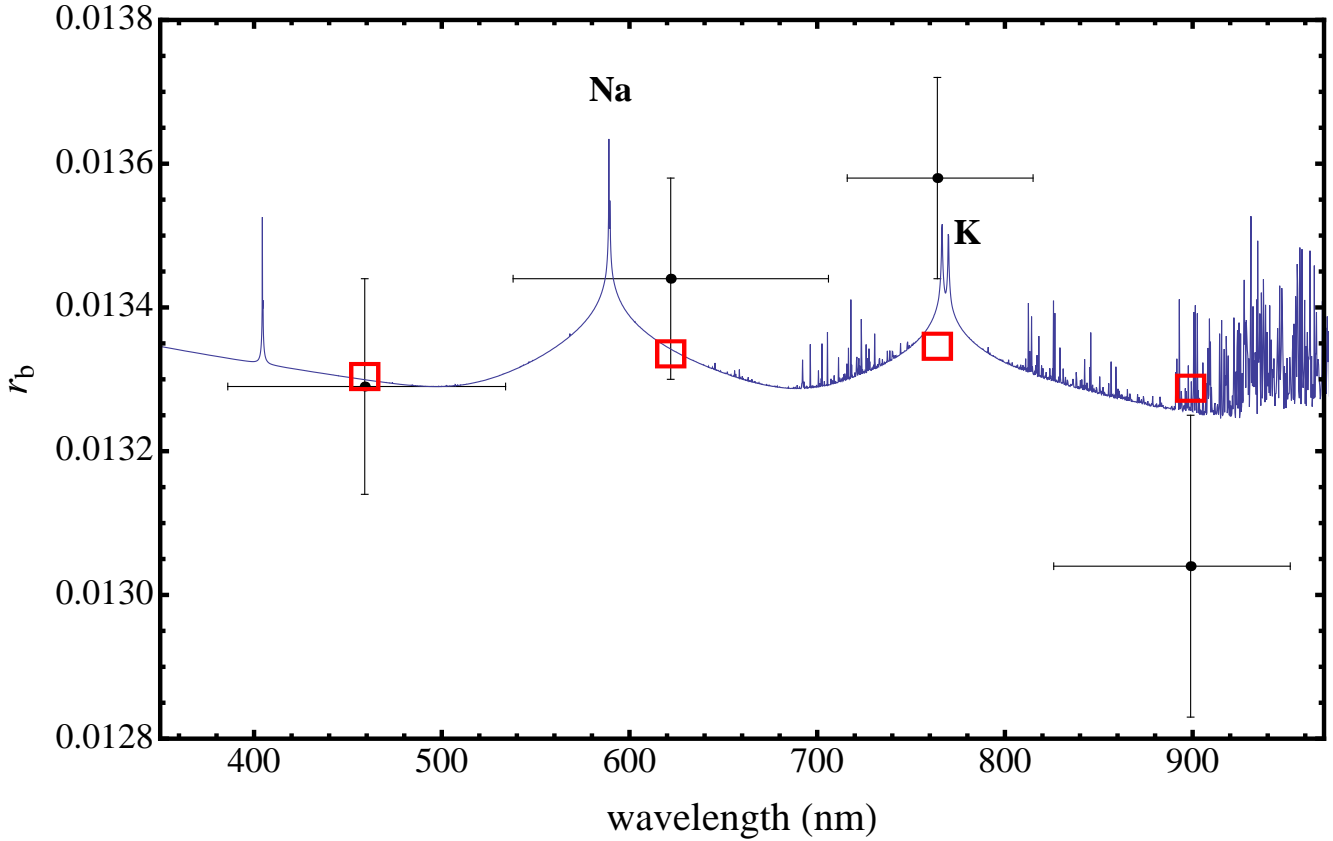}}%
\caption{As in Fig.\,\ref{Fig_5} but zoomed in on the optical wavelengths. In this range, the synthetic transmission spectrum is dominated by gaseous Na and K.}
\label{Fig_6}
\end{figure*}

At optical wavelengths, $r_{\mathrm{b}}$ appears to be slightly
larger in $r^\prime$ and $i^\prime$ than $g^\prime$ and
$z^\prime$. This is as expected for a planet with an equilibrium
temperature $T_{\rm eq}\approx 1300$\,K, due to the larger
predicted radius around the Na and K lines. Although a similar
effect is seen in the model (Figure\,\ref{Fig_6}), our data are
not sufficiently precise to reject the default hypothesis that
$r_{\mathrm{b}}$ is constant with wavelength. The radius of
WASP-44\,b is in agreement with the theoretical model in the $J$
band, but is larger in the $H$ and $K$ bands. We found that the
$r_\mathrm{b}$ in $r^\prime$ is $9.5\%$ and $4.7\%$ smaller than
that for the $H$ and $K$ band respectively, which correspond to a
difference in units of atmospheric pressure scale height ($H$) of
roughly 23$H$ and 10$H$, respectively. The first difference is
very large, and would involve an extreme opacity in $H$.

A more plausible alternative is that the NIR photometry is affected by systematics or correlated noise, which was not possible to eliminate. This is also suggested by the poor SNR observed in the data, which is reflected in the higher scatter of the points in the $J,\,H,\,K$ light curves and the different shapes of the fit in the same band between the first and the second observed transits (Figs.\ \ref{Fig_2} and \ref{Fig_3}). We deduce that systematic errors in the NIR dominate the error budget at these wavelengths.

\section{Summary and discussion} %
\label{Sec_4}

In this work we have presented the first photometric follow-up of the transiting extrasolar planetary system WASP-44. One transit was observed using WFC on the INT, and two transits were observed in seven passbands simultaneously using GROND on the MPG/ESO 2.2\,m telescope. Using these new datasets plus another one taken from the literature \citep{anderson2011}, we have obtained improved measurements of the orbital ephemerides and physical properties of the system. In particular, we found that the radius of WASP-44\,b is smaller than previously found, and is different (at $2\sigma$ confidence level) from that theoretically expected for a free-core planet \citep{fortney2007}. This suggests that WASP-44\,b is a ``heavy element rich'' planet, with the heavy elements being in a distinct core or mixed within the H/He envelope. Such a result is important in order to draw an accurate \emph{mass--radius} plot for TEPs, which provides key diagnostic for theoretical works that look to infer the bulk composition of the giant planets and distinguish them (in some cases) from brown dwarfs.

Although GROND was not designed for follow-up studies of planetary
transits, it is one of the very few imaging systems worldwide able
to perform simultaneous photometric multi-band observations. The
later fact makes GROND an efficient and very useful instrument to
detect transit anomalies or to probe the atmosphere of TEPs.
Indeed, atomic and molecular absorption as well as scattering
processes may result in detectable radius variations as a function
of wavelength \citep{fortney2007,batygin2009}. \citet{demooij2012}
already used GROND in the case of GJ\,1214\,b, but they restricted
their analysis only to the optical bands. Here, for the first
time, we used all the seven band of this powerful instrument, to
search for variations of the radius of WASP-44\,b with wavelength.
We measured the planetary radius in the seven GROND passbands,
corresponding to a wavelength coverage of 370--2440\,nm, and
compared them to a synthetic spectrum based on an isothermal model
atmosphere in chemical equilibrium. The model reproduces the
radius values in the optical, in agreement with the equilibrium
temperature of WASP-44\,b but, our data lack the precision to rule
out the possibility that the planetary radius does not vary with
wavelength. It does not predict the radius values found in the $H$
and $K$ bands, which differ by roughly 23 and 10 atmospheric
pressure scale heights to those in the other bands. This
phenomenon is most likely due to the comparatively low quality of
the NIR data. We have worked to reduce the sources of noise in the
subsequent uses of the GROND instrument. More enticing results
were obtained for other TEPs and will be shown in forthcoming
papers.

\section*{Acknowledgments}

Based on observations collected at the MPG/ESO 2.2\,m Telescope located at ESO La Silla, Chile. Operations of this telescope are jointly performed by the Max Planck Gesellschaft and the European Southern Observatory. GROND has been built by the high-energy group of MPE in collaboration with the LSW Tautenburg and ESO, and is operating as a PI-instrument at the MPG/ESO 2.2\,m telescope. We thank David Anderson for supplying photometric data, and Timo Anguita and R\'{e}gis Lachaume for their technical assistance during the observations. We thank the anonymous referee for their useful criticisms and suggestions that helped us to improve the quality of the present paper. The reduced light curves presented in this work will be made available at the CDS (http://cdsweb.u-strasbg.fr/). JS acknowledges financial support from STFC in the form of an Advanced Fellowship.

\appendix

\section{GROND-NIR data Reduction}
\label{Appendix_A}

The calibration of the GROND NIR data is a little more complicated
than the optical data, because of the presence of electronic
odd-even read-out pattern along the y-axis. In order to remove
this pattern, we smooth each image and compare it with the
unsmoothed ons after the master dark has been subtracted. The
amplitudes of the read-out pattern are then determined by
comparing the median level of each column to the overall median
level. Each column is shifted back to the overall median level.
The master skyflats are divided out from the science images after
the removal of read-out pattern.

Since our observations are in staring mode, sky background is an
important contribution in the images. With the 20-position
dithering sky measurements right before and after the science time
series, we could construct a sky emission model for each science
image. These sky images are calibrated in the same way as the
science images. We mask out all the stars in the calibrated sky
images, and normalise each image with their own overall median
levels. All the normalised skies are then median combined, and
normalised again. For each science image, we scale the pre- and
post-science sky models to their background individually. The
final sky model is constructed by weighted combining them
according to the inverse square of fitted $\chi^2$. This sky model
is then subtracted from corresponding science image. After sky
subtraction, we still see remnant structure, which is expected due
to the variation of the sky during the observing sequence.

After these calibrations have been applied, we perform aperture
photometry on WASP-44 as well as three nearby comparison stars.
The locations of each star are determined by {\sc idl/find}, which
calculates the centroids by fitting Gaussians to the marginal x
and y distributions. In order to find the optimal photometry, we
lay 30 apertures on all the stars in step of 0.5 pixel, each with
ten annuli in step of 1 pixel. This produces 300 datasets with
different aperture settings. In this way, the above mentioned sky
subtraction remnant effect is accounted for in the sky annulus
adopted in the aperture photometry. We divide the flux of each
star with its own median value. A composite reference light curve
is constructed by median-combining fluxes of ensembles of
comparison stars, which are required to show the least deviation
from the target. We normalixe the target light curve with this
composite reference light curve.

Since the normalised light curve still shows strong red noise
correlated with positions, seeing and airmass, instead of
selecting the optimal photometry directly from $rms$ of
out-of-transit baseline, we fit the whole light curve
simultaneously with the theoretical light curve model multiplied
with a baseline correction function. This correction function
consists of positions/time in quadratic form, and FWHMs/airmass in
linear form. We tried to select the optimal light curves by
choosing the ones with the least $rms$ of O--C flux residuals
among all the datasets. However, the transit depths of the optimal
light curves of the same band from different nights are very
deviant from each other. We note that light curves of the same
band obtained using different aperture and annulus sizes could
exhibit similar least $rms$, however produce different transit
depths. Therefore, for each NIR band, we choose the two light
curves with the least $rms$ and with consistent transit depth
between the two nights as our final optimal photometric results.

\label{lastpage}

\end{document}